\documentclass[aps,prc,showpacs,nofootinbib,twocolumn]{revtex4-1}

\usepackage{epsfig}  
\vspace*{1.0cm}
\usepackage{graphicx,amsmath}
\usepackage{color}
\newcommand\ba{\begin{eqnarray}}
\newcommand\ea{\end{eqnarray}}

\newcommand{\be}{\begin{equation}}
\newcommand{\ee}{\end{equation}}
\newcommand{\bas}{\begin{eqnarray*}}
\newcommand{\eas}{\end{eqnarray*}}
\begin{document}
\vspace*{-1.5cm}
\title{Variation of hadronic and nuclei mass level oscillation periods for different spins.}

\author{B. Tatischeff}
\email{tati@ipno.in2p3.fr}
\affiliation{CNRS/IN2P3, Institut de Physique Nucl\'eaire, UMR 8608, Orsay, F-91405\\
 and Univ. Paris-Sud, Orsay, F-91405, France\\
}
\vspace*{-1.5cm}
\begin{abstract}
A systematic study of hadronic masses shows regular oscillations that can be fitted by a simple cosine function. This property can be observed when  the difference between adjacent masses of each family is plotted versus the mean mass. This symmetry of oscillation is also observed for the nuclear level masses of given spin. 
\keywords{Hadronic masses, mesons baryons,symmetry}
\end{abstract}
\maketitle

\section{Introduction}

A new property of particle masses was recently shown when studying the mass variation versus the mass increase for adjacent meson and baryon masses of given families \cite{boris}. 
The investigated function is:
\be 
 m_{(n+1)} - m_{n} = f [(m_{(n+1)} + m_{n})/2]\\
\ee
where  $m_{(n+1)}$ corresponds to the (n+1) hadron mass value. The difference of two successive masses was plotted versus the mean  value of the two nearby masses. Such studies were restricted to hadron families holding at least five masses. Regular oscillations were observed 
giving rise to a new symmetry, the symmetry of oscillation. It was noticed in \cite{boris} that "the existence of composite hadrons, results from the addition of several forces, related to strong interaction, that combine in, at least, one attractive and one repulsive force. The equilibrium among these forces allows the hadron to exist, otherwise the composite mass will either disintegrate, or mix into a totally new object with loss of the individual components". As in classical physics, these opposite forces may generate oscillating behaviour. 

The obtained data are fitted using a cosine function:
\be
\Delta M = \alpha_{0} + \alpha_{1} cos( (M - M_{0}) / M_{1}) \\
\ee
where M$_{0}$ /M$_{1}$ is defined within 2$\pi$. All coefficients, and masses used to draw the figures are in MeV units. The quantitative information is given in Tables I and II presented below. The oscillation periods are P = 2 $\pi  M_{1}$. Both $\alpha_{0}$ and $\alpha_{1}$ are adjusted on the extreme values on all figures.

Whereas smaller periods than those given in the Tables may also reproduce the data, we show in the following figures, the largest possible values. 

The discussion concerned the oscillatory periods, and not the oscillation amplitudes which need theoretical study outside the scope of previous and present papers.

For the same reason, the existence of substructures in hadrons, we expect to observe oscillations in nuclei made with nucleons. Such study will be considered after the hadronic masses. 

\section{Application to hadronic masses}
The masses and widths are read from the Review of Particle Physics \cite{olive}, taking into account all the data reported, even if, in some cases, they are omitted from the summary table. 

In our previous paper \cite{boris} data were shown for the following meson families with fixed quantum numbers:  $f_0$ in fig. 1(a), $f_2$ in fig. 1(b), 
charmonium (${\it c}{\bar c}$) $0^{-}(1^{--})$ in fig. 4(a) and  bottomonium (${\it b}{\bar b})$
$0^{-}(1^{--})$ in fig. 4(b). 
Several other data were also shown in the same paper without restriction to given quantum numbers for charmed in fig. 2(a), charmed strange in fig. 2(b), (${\it c}{\bar c}$) in fig. 3(a), (${\it b}{\bar b}$) in fig. 3(b) mesons. The corresponding figures displayed several data outside the fitting curves, mainly for  (${\it c}{\bar c}$)  but also for charmed strange mesons.

Resulting data were also shown for several baryon families without selection of given quantum numbers, contrary to the indication reported in column $J^{PC}$ of \cite{boris}. They are: $N^{*}$ in fig. 5(a), $\Delta$ in fig. 6(a), $\Lambda$ in fig. 6(b), $\Lambda_C$ in fig. 5(b),  $\Xi$ in fig. 7(a) and $\Xi_C$ in fig. 7(b).  

 All previous data were fitted with oscillations with use of a few first masses. The fit gets often spoiled over a few MeV.  The comparison between the selection of charmonium and bottomonium data with fixed quantum numbers, and without spin selection, suggests the relevance  to restrict  the study to particle families with given spins. Of course the necessity to have still at least five known masses remains. This condition will reduce the possibilities of application.                
 
This study is  done below, where the figures shown in \cite{boris} with such criteria of given spin, are repeated here in  purpose of consistency.
The masses reported in \cite{olive} are used independently of the number of 
attributed stars. When the name is different from the mass, the mass is used.

\begin{figure}[ht]
\centering
\hspace*{-0.2cm}
\scalebox{0.58}[0.68]{
\includegraphics[bb=19 136 537 546,clip,scale=0.8] {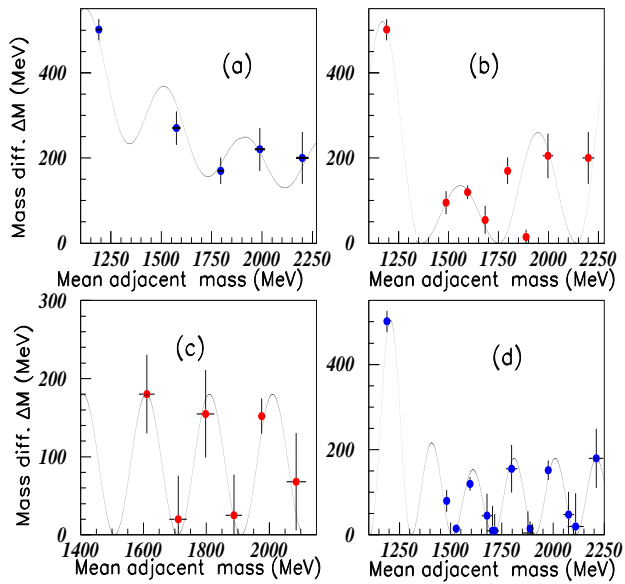}}
\caption{Color on line. Inserts (a), (b), (c), and (d) show successively the mass difference between successive masses, plotted versus both corresponding mean masses for
$N^{*}$ (1/2$^{+}$),
$N^{*}$ (1/2$^{+}$ + 1/2$^{-}$), $N^{*}$ (3/2$^{+}$ + 3/2$^{-}$), and
$N^{*}$ (1/2$^{+}$ +1/2$^{-}$ + 3/2$^{+}$ + 3/2$^{-}$)  baryons.}
\end{figure}

Fig. 1 shows in inserts (a), (b), (c), and (d) the data for $N^{*}$ (1/2$^{+}$), 
$N^{*}$ (1/2$^{+}$ + 1/2$^{-}$), $N^{*}$ (3/2$^{+}$ + 3/2$^{-}$), and
$N^{*}$ (1/2$^{+}$ +1/2$^{-}$ + 3/2$^{+}$ + 3/2$^{-}$). Although two data in insert (b) lie outside the curve, both fits in inserts (a) and (b) are obtained with the same period P = 390~MeV. We will use that to add subsequently the masses having the same spin but different parities, and allow therefore to get more data with five, or more masses analysed simultaneously. 
Fig. 1(c) shows nice fit for 
$N^{*} (3/2^{+}$) +  $N^{*} (3/2^{-})$. The period here is P =201~MeV. Fig. 1(d) shows the result for $N^{*}$ baryon masses, having both parities and both spins (1/2) and (3/2). These data are fitted with P = 201~MeV. 

Therefore we will later on add the data having different parities and study them separately for different spins.
\begin{figure}[ht]
\centering
\hspace*{-0.2cm}
\scalebox{0.58}[0.68]{
\includegraphics[bb=21 331 520 546,clip,scale=0.8] {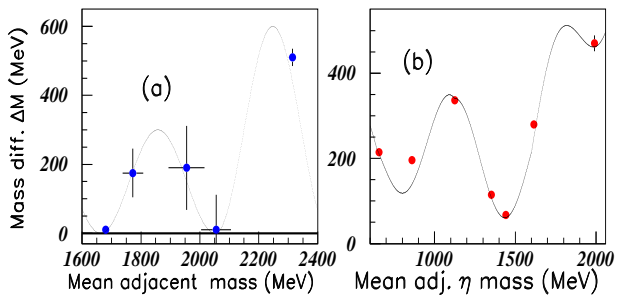}}
\caption{Color on line. Inserts (a) and (b) show successively the mass difference between successive masses, plotted versus both corresponding mean masses for
$N^{*}$ (5/2$^{+}$ + 5/2$^{-}$) baryons and $\eta  (0^{-})$ mesons.}
\end{figure}
\begin{figure}[b]
\centering
\scalebox{0.58}[0.68]{
\includegraphics[bb=17 336 524 551,clip,scale=0.8] {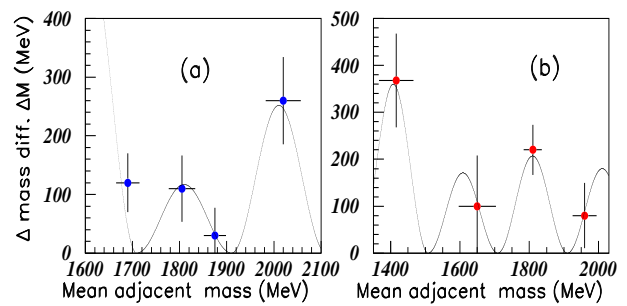}}
\caption{Color on line. Inserts (a) and (b) show successively the mass difference between successive masses, plotted versus both corresponding mean masses for
$\Delta$ (1/2),  $\Delta$ (3/2) baryons.}
\end{figure}
\begin{figure}[t]
\centering
\hspace*{-0.2cm}
\scalebox{0.58}[0.68]{
\includegraphics[bb=20 137 518 551,clip,scale=0.8] {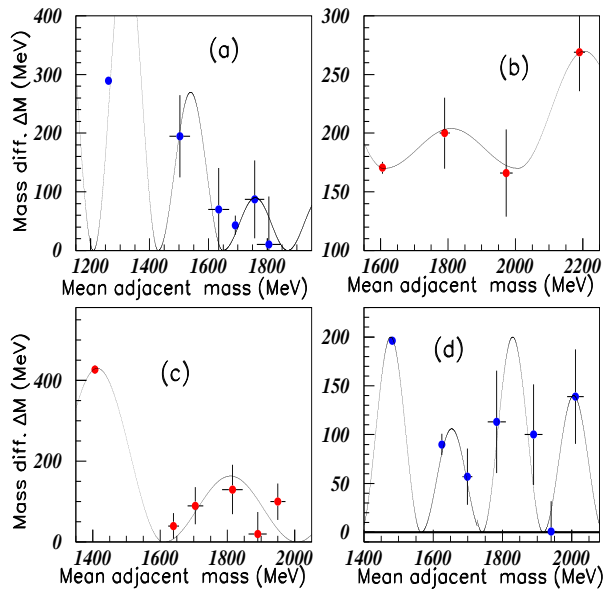}}
\caption{Color on line. Inserts (a),  (b), (c), and (d) show successively the mass difference between successive masses, plotted versus both corresponding mean masses for $\Lambda~(1/2^{+})$ and ($1/2^{-}$), $\Lambda~(3/2^{+})$ and ($3/2^{-})$, $\Sigma~(1/2^{+})$ and ($1/2^{-}$), $\Sigma~(3/2^{+})$ and $(3/2^{-})$ baryons.}   
\end{figure}
\begin{figure}[ht]
\centering
\hspace*{-0.2cm}
\scalebox{0.58}[0.68]{
\includegraphics[bb=13 131 522 549,clip,scale=0.8] {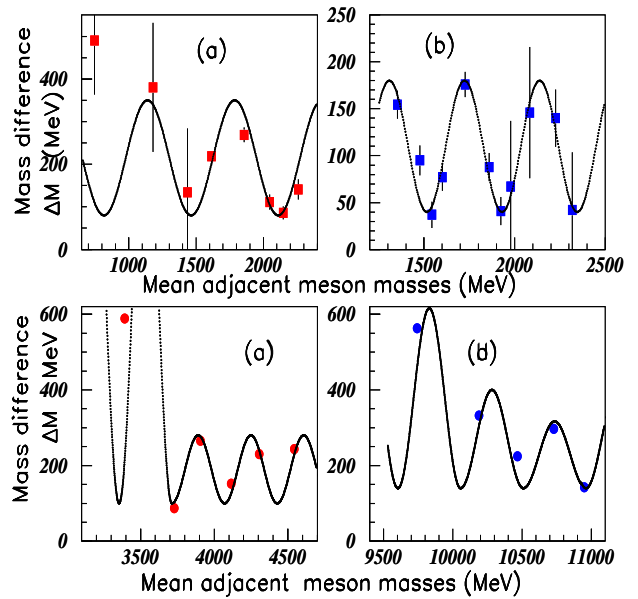}}
\caption{Color on line. Inserts (a), (b), (c), and (d) show successively the mass difference between successive masses, plotted versus both corresponding mean masses for $f_0$ ($0^{++}$),   $f_2$ ($2^{++}$), (${\it c}{\bar c}$) $0^{-}(1^{--})$, and  (${\it b}{\bar b})$ $0^{-}(1^{--})$ mesons.}
\end{figure}
\begin{figure}[ht]
\centering
\hspace*{-0.2cm}
\scalebox{0.58}[0.68]{
\includegraphics[bb=24 136 525 552,clip,scale=0.8] {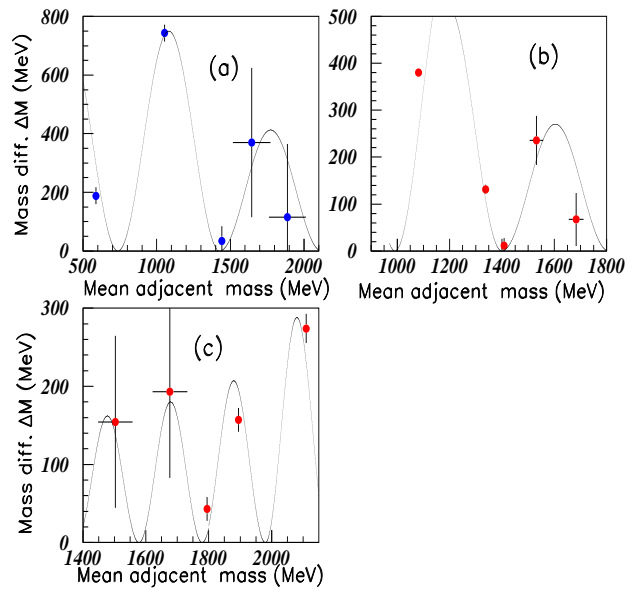}}
\caption{Color on line. Inserts (a), (b), and (c),  show successively the mass difference between successive masses, plotted versus both corresponding mean masses for K ($0^{+}$) and K ($0^{-}$),  K ($1^{+}$) and K ($1^{-}$), and K ($2^{+}$) and K ($2^{-}$)  mesons.}
\end{figure}
\begin{figure}[h]
\centering
\hspace*{-0.2cm}
\scalebox{0.58}[0.68]{
\includegraphics[bb=18 231  518 547,clip,scale=0.8] {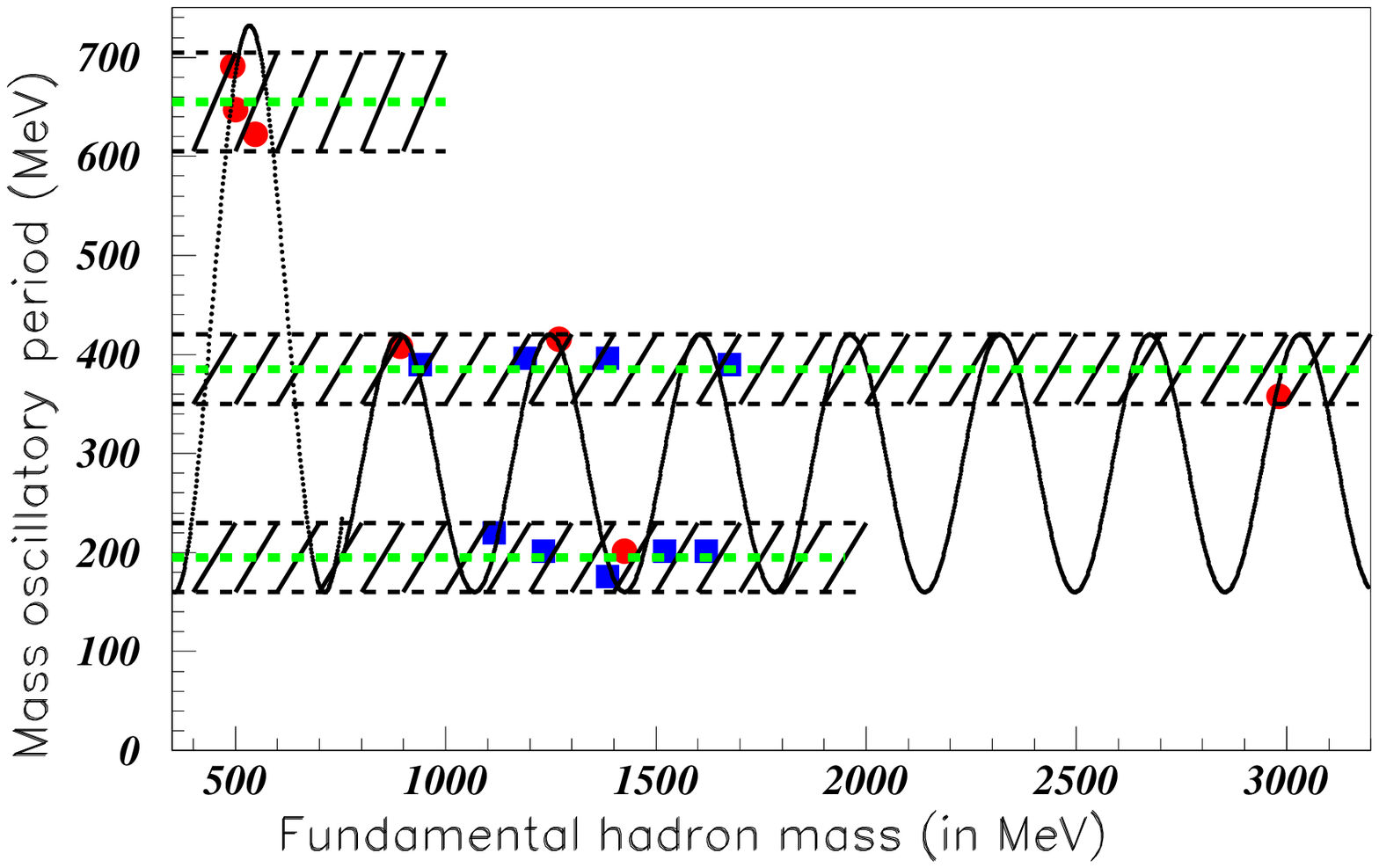}}
\caption{Color on line. Variation of the oscillating periods of the hadronic mass families. Full red circles (blue squares) show the results for mesons (baryons).}
\end{figure}

Fig. 2 shows in insert (a) the mass difference between successive masses, plotted versus both corresponding mean masses for $N^{*}$ (5/2$^{+}$) + $N^{*}$ (5/2$^{-}$). The corresponding period 
P~=~390~MeV is the same as the one obtained to describe the variation of the data for $N^{*}$~(J=1/2) fig.~1(b) and $N^{*}$~(J=3/2) fig.~1(c) baryons.

The data and fit for the $\eta$ meson are shown in fig.~2(b). The very precised masses are well fitted  with the period P = 622~MeV. A mass at M = 762~MeV is introduced, following the recent suggestion \cite{BTETG}.

Fig. 3 shows the data for $\Delta$ baryons J = (1/2) in insert (a) and J = (3/2) in insert (b), without distinguishing the parities. Both data are fitted with the same period 
P~=~201~MeV.
  
Fig. 4 shows the the data for strange baryons $\Lambda~(1/2^{+})$ and ($1/2^{-}$) in insert (a), $\Lambda~(3/2^{+})$ and ($3/2^{-}$) in insert (b),
$\Sigma (1/2^{+})$ and $(1/2^{-})$ in insert (c), $\Sigma (3/2^{+})$ and $(3/2^{-})$ in insert (d). The data in fig. 4(a) are fitted by an oscillating curve, although here more simple functions are possible.

Fig. 5 shows the results for  $f_{0}$ ($0^{++}$) and $f_{2}$ ($2^{++}$) light unflavoured mesons in inserts (a) and  (b). 
 The $\sigma$ or $f_{0}$(500) meson is broad and its mass (which unprecision is taken to $\Delta$M = 125~MeV), is badly determined. The reasonable fit allows to extrapolate the masses of the next $f_{0}$ not extracted experimentally up to now.  They are  M $\approx$ 2670 and 2760~ MeV. In the same way, the masses of the next $f_{2}$ mesons can be tentatively predicted to be: M $\approx$ 2380, 2450, and 2625 MeV. 
 Inserts (c) and (d) show the results for 
(${\it c}{\bar c}$) $0^{-}(1^{--})$, and  (${\it b}{\bar b})$ $0^{-}(1^{--})$ mesons.
The mass of the last quoted meson used in fig. 5(c), X(4660) $?^{?}(1^{--})$ fits perfectly in this  distribution, and is therefore kept,  assigning tentatively the same quantum numbers.   The extrapolation allows to predict tentatively the next corresponding $\Psi$  masses: M $\approx$ 4805 and 5080~MeV.  In the same way, the tentatively extrapolated $\Upsilon$ masses are: M $\approx$ 11330 and 11560~MeV. 

Fig. 6 shows the data for strange kaons, K ($0^{+ -}$) in insert (a),  
K ($1^{+ -}$) in insert (b), and K ($2^{+ -}$) in insert (c). Oscillatory shapes must be used for fits. The corresponding periods decrease regularly from P = 691~MeV, to P=408~MeV, and finally to P=201~MeV for increasing spins in inserts: (a), (b), and (c).  

\begin{table}[t]
\caption{Quantitative information concerning the oscillation behavior of some mesons and baryons analysed previously. P is the period (in MeV).}
\label{Table I}
\vspace{5.mm}
\begin{tabular}{c c c c c c}
\hline\underline{}
name&q.c.&fig.&J&mass&P\\
\hline
$K_{0}$&$q{\bar s}$&6(a)&0&493.7&691\\
$\eta$&$q{\bar q}$&2(b)&0$^{-+}$&547.9&622\\
$f_{0}$&$q{\bar q}$&5(a)&$0^{++}$&475&647\\
$K_{1}$&$q{\bar s}$&6(b)&1&892&408\\
$f_{2}$&$q{\bar q}$&5(b)&$2^{++}$&1275&415\\
$K_{2}$&$q{\bar s}$&6(c)&2&1425&201\\
charm.&$c{\bar c}$&5(c)&$1^{--}$&2981.5&358\\
botto.&$b{\bar b}$&5(d)&$1^{--}$&9391&452\\
$N^{*}$&qqq&1(b)&1/2&939&390\\
$N^{*}$&qqq&1(c)&3/2&1520&201\\
$N^{*}$&qqq&2(a)&5/2&1675&390\\
$\Lambda$&qqs&4(a)&1/2&1115.7&220\\
$\Lambda$&qqs&4(b)&3/2&1519.5&396\\
$\Delta$&qqq&3(a)&1/2&1620&201\\
$\Delta$&qqq&3(b)&3/2&1232&201\\
$\Sigma$&qqs&4(c)&1/2&1189.4&396\\
$\Sigma$&qqs&4(d)&3/2&1385&176\\
\hline
\end{tabular}
\end{table}

All obtained periods fitting the previous data are reported in table I and fig.~7. Meson (baryon) periods are plotted with full red circles (full blue squares) versus the lower mass of each family.  They all are located in three well defined ranges, the same for mesons and baryons. Three general properties are observed:

The periods of the three (meson) families with the lowest spin (J = 0):
$f_{0}$, $K^{0}$, and $\eta$ are located in the highest range close to P$\approx$655~MeV.

The periods corresponding to the other families  distributed in the two other ranges,
 favor the intermediate range for lower spin. This is true for periods of K ~(J=1) compared to K~(J=2), N~$(J=1/2)$ compared to N~$(J=3/2)$,  $\Sigma$~$(J=1/2)$ compared to $\Sigma~$(J=3/2). However the opposite is observed for periods corresponding to 
$\Lambda~(J=1/2)$ and $\Lambda~(J=3/2)$. And also the period of oscillation corresponding to the masses of N (J=5/2) is larger than that of N (J=3/2) and is equal to that of N (J=1/2), suggesting here again an oscillatory behaviour. Such behaviour is indeed observed in fig.~7 with P~=~357~MeV, better adjusted to meson than to baryon results. So the periods of $\Delta~(J=1/2)$ and $\Delta~(J=3/2)$) lie outside the distribution. 

We notice that the distribution reported in fig.~7, fits the period correponding to (${\it c}{\bar c}$) $0^{-}(1^{--})$ mesons and also the period corresponding to (${\it b}{\bar b})$ $0^{-}(1^{--})$ mesons not plotted on fig.~7 since the very large gap between masses. The distribution reported in fig.~7 differs from that reported in \cite{Tatischeff:2016psc}. New data are analysed in the present paper, when several data, without spin selection,  were used in \cite{Tatischeff:2016psc}.

The mean values of the three ranges shown in fig.~7, are pointed out by dashed  lines (in green on line) at M(0) $\approx$ 655~MeV, M(1) $\approx$ 385~MeV, and M(2) $\approx$ 195~MeV. The meson oscillating periods, shown in red full circles, are gathered together following the relation: J - $\left |{S}\right |$ + 2*I  which value equals 0, 1, and 2 for the three ranges from up to down. This relation does not apply for ${\it f_{2}}$ mesons.

\section{Application to nuclei masses}
\begin{figure}[h]
\centering
\hspace*{-0.2cm}
\scalebox{0.58}[0.68]{
\includegraphics[bb=39 136 518 546,clip,scale=0.8] {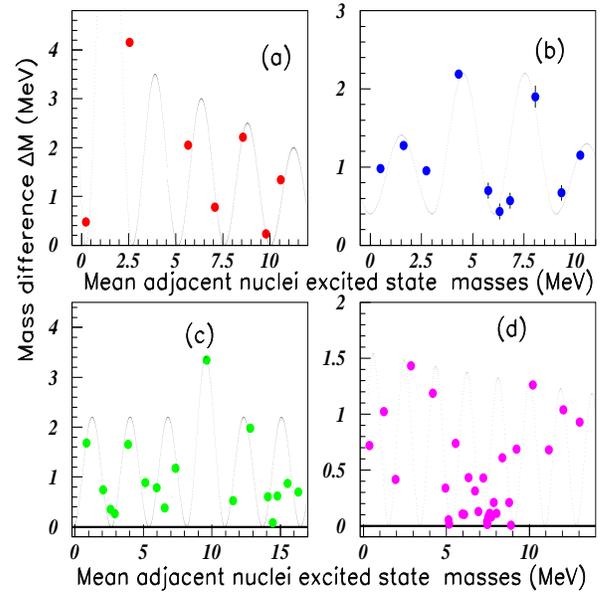}}
\caption{Color on line. Inserts (a), (b), (c), and (d) show successively the mass difference between successive masses, plotted versus both corresponding mean masses for respectively $^{7}$Li, $^{8}$Li, $^{9}$Be, and $^{10}$B nuclei.}
\end{figure}
\begin{figure}[h]
\centering
\hspace*{-0.2cm}
\scalebox{0.58}[0.90]{
\includegraphics[bb=28 229 532 546,clip,scale=0.8] {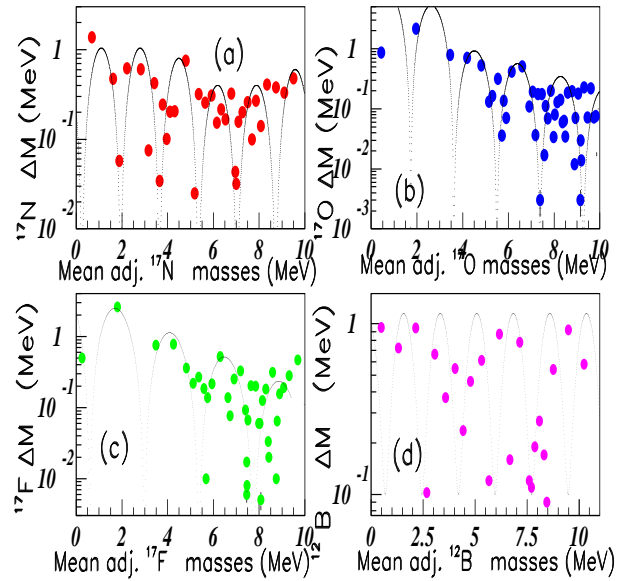}}
\caption{Color on line. Inserts (a), (b), (c), and (d) show successively the mass difference between successive masses, plotted versus both corresponding mean masses for   respectively $^{17}$N, $^{17}$O, $^{17}$F, and $^{12}$B nuclei.}
\end{figure}

It is reasonable to expect oscillations in the nuclei mass levels for the same reason as before for hadrons. The nucleons in nuclei are bound by opposite forces. This property is studied below using data from \cite{Lederer} \cite{Ajzenberg-Selove:1979uqf,Ajzenberg-Selove:1968ztr,Ajzenberg-Selove:1976uid} when not specified. We start the analysis without spin selection, considering all level masses.

Fig.~8 shows the mass difference between successive masses plotted versus the corresponding massses for $^{7}$Li (red) P=2.45~MeV, $^{8}$Li (blue) P=3.02~MeV, $^{9}$Be (green) P=2.76~MeV, and $^{10}$B (purple) P=1.88~MeV respectively in inserts (a), (b), (c), and (d). We observe an increase of the level number with increasing mass. We observe also that the fit between data and calculated curves spoils after the five-six first MeV in the case of $^{10}$B nucleus.
For heavier nuclei, these properties are amplified as seen in fig.~9.

Fig.~9 shows the mass difference between successive masses plotted versus the corresponding masses for $^{17}$N P=1.70~MeV, $^{17}$O P=1.88~MeV, $^{17}$F 
P=2.39~MeV, and $^{12}$B P=1.76~MeV nuclei. In spite of the large mass differences for all data, emphasized by the log scale, the first data are rather well fitted, then followed by a large number of spread data.  

So the situation is comparable to the one observed for hadrons, and therefore brings us to separate the nuclei level masses by their spins.

The next figures will study the oscillation properties of nuclei level masses having the same spin.
Although a large number of level masses are known for the majority of nuclei, rather few have a number of known quantum numbers allowing the same studies as previously done (five or more level masses with the same spin).     

\begin{figure}[h]
\centering
\hspace*{-0.2cm}
\scalebox{0.58}[0.68]{
\includegraphics[bb=20 136 515 554,clip,scale=0.8] {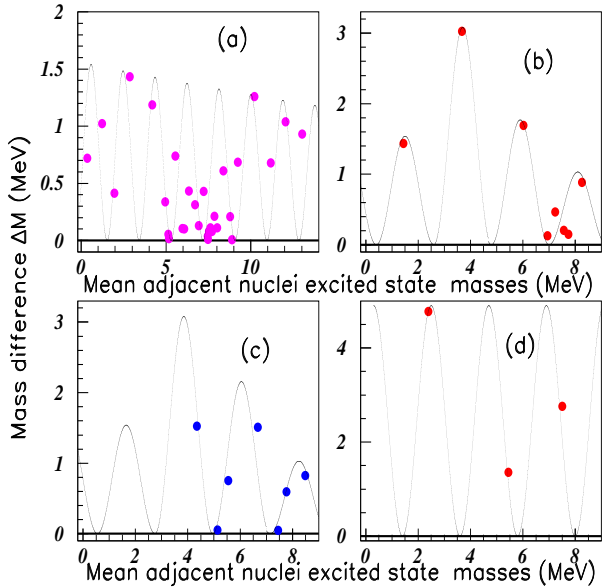}}
\caption{Color on line. Inserts (a), (b), (c), and (d) show successively the mass difference between successive masses, plotted versus both corresponding mean masses for $^{10}$B nucleus: all spins, J=1, J=2, and J=3  respectively.}
\end{figure}

\begin{figure}[h]
\centering
\hspace*{-0.2cm}
\scalebox{0.58}[0.68]{
\includegraphics[bb=20 136 515 554,clip,scale=0.8] {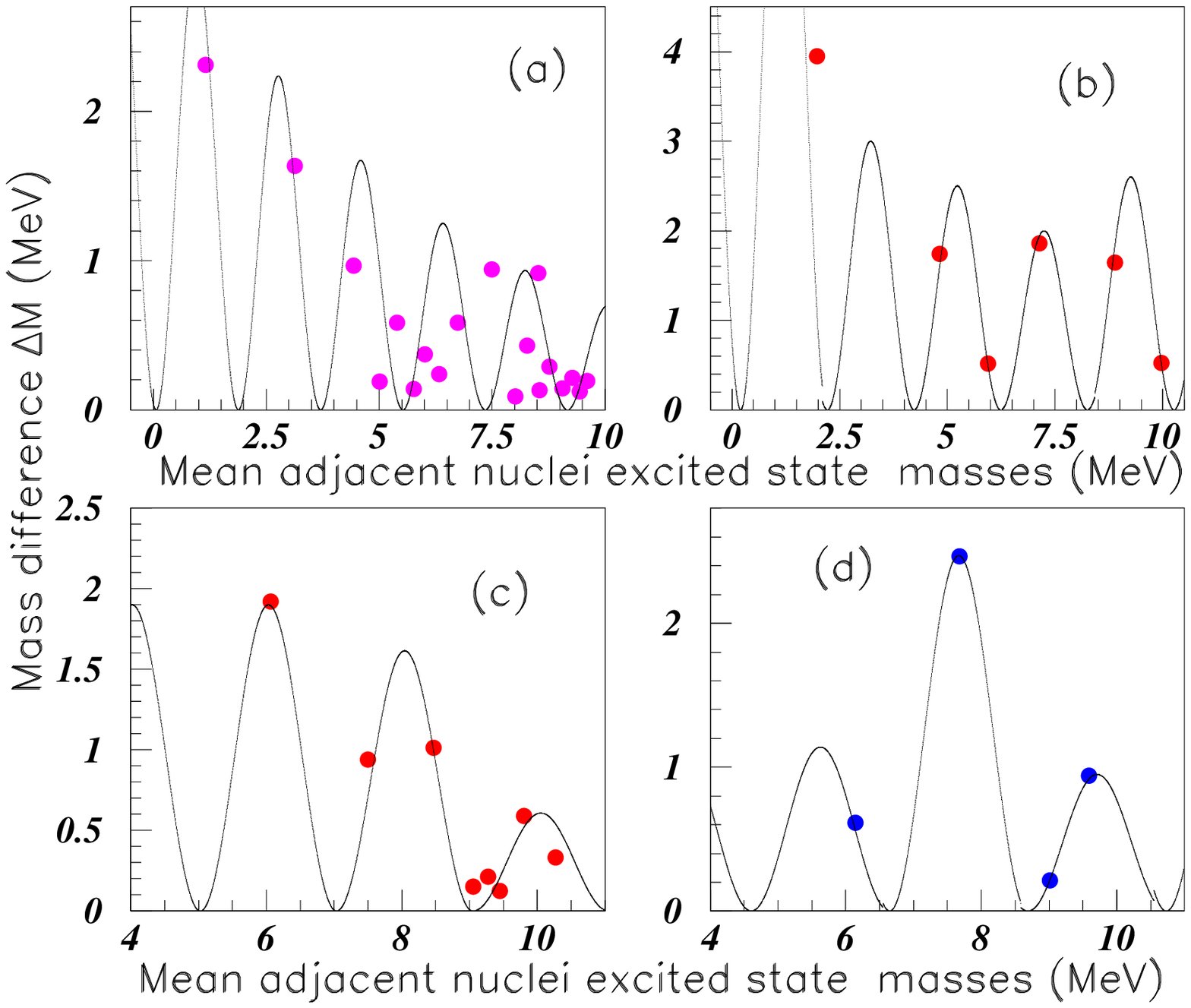}}
\caption{Color on line. Inserts (a), (b), (c), and (d) show successively the mass difference between successive masses, plotted versus both corresponding mean masses for $^{14}$N nucleus: all spins, J=1, J=2, and J=3 respectively.}
\end{figure}

\begin{figure}[h]
\centering
\hspace*{-0.2cm}
\scalebox{0.58}[0.68]{
\includegraphics[bb=40 138 520 547,clip,scale=0.8] {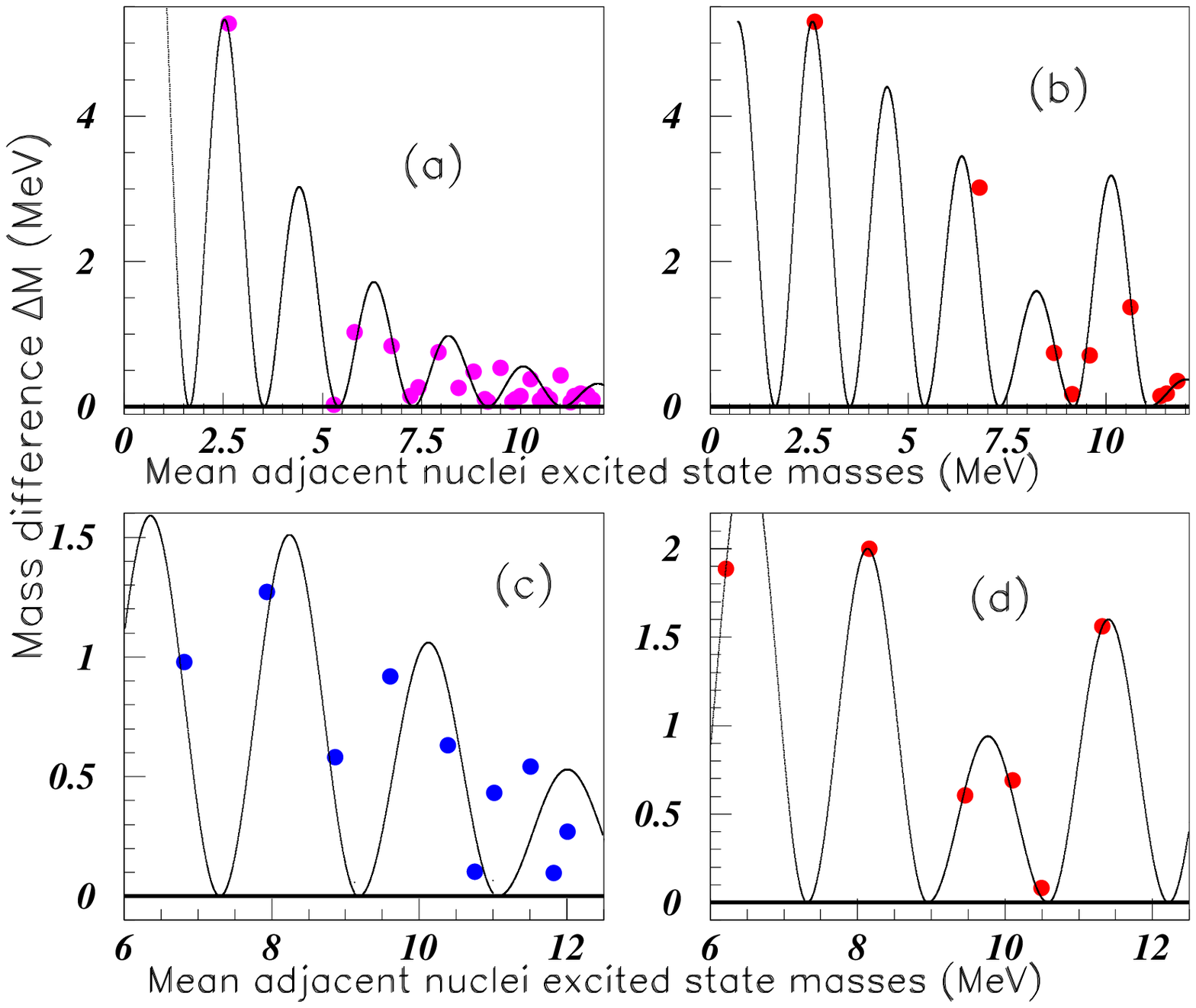}}
\caption{Color on line. Inserts (a), (b), (c), and (d) show successively the mass difference between successive masses, plotted versus both corresponding mean masses for $^{15}$N nucleus: all spins, J=1/2, J=3/2, and J=5/2 respectively.}
\end{figure}

\begin{figure}[h]
\centering
\hspace*{-0.2cm}
\scalebox{0.58}[0.68]{
\includegraphics[bb=22 238 517 546,clip,scale=0.8] {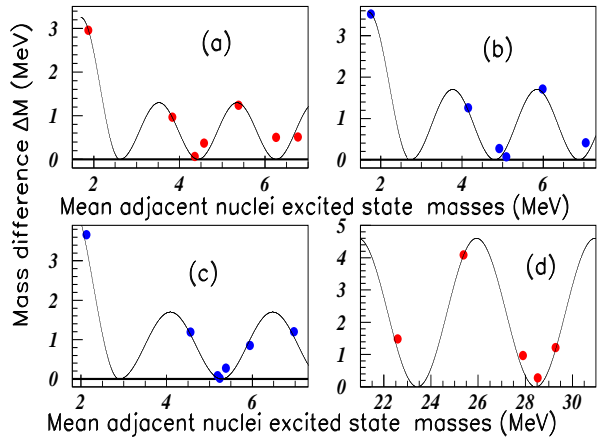}}
\caption{Color on line. Inserts (a), (b), and (c) show successively the mass difference between successive masses, plotted versus both corresponding mean masses for $^{16}$N nucleus: J=1, J=2, and J=3 respectively. Insert (d) shows the result for $^{4}$He J=2 levels, P=5.03~MeV.}
\end{figure}

\begin{figure}[h]
\centering
\hspace*{-0.2cm}
\scalebox{0.58}[0.68]{
\includegraphics[bb=21 83 518 546,clip,scale=0.8] {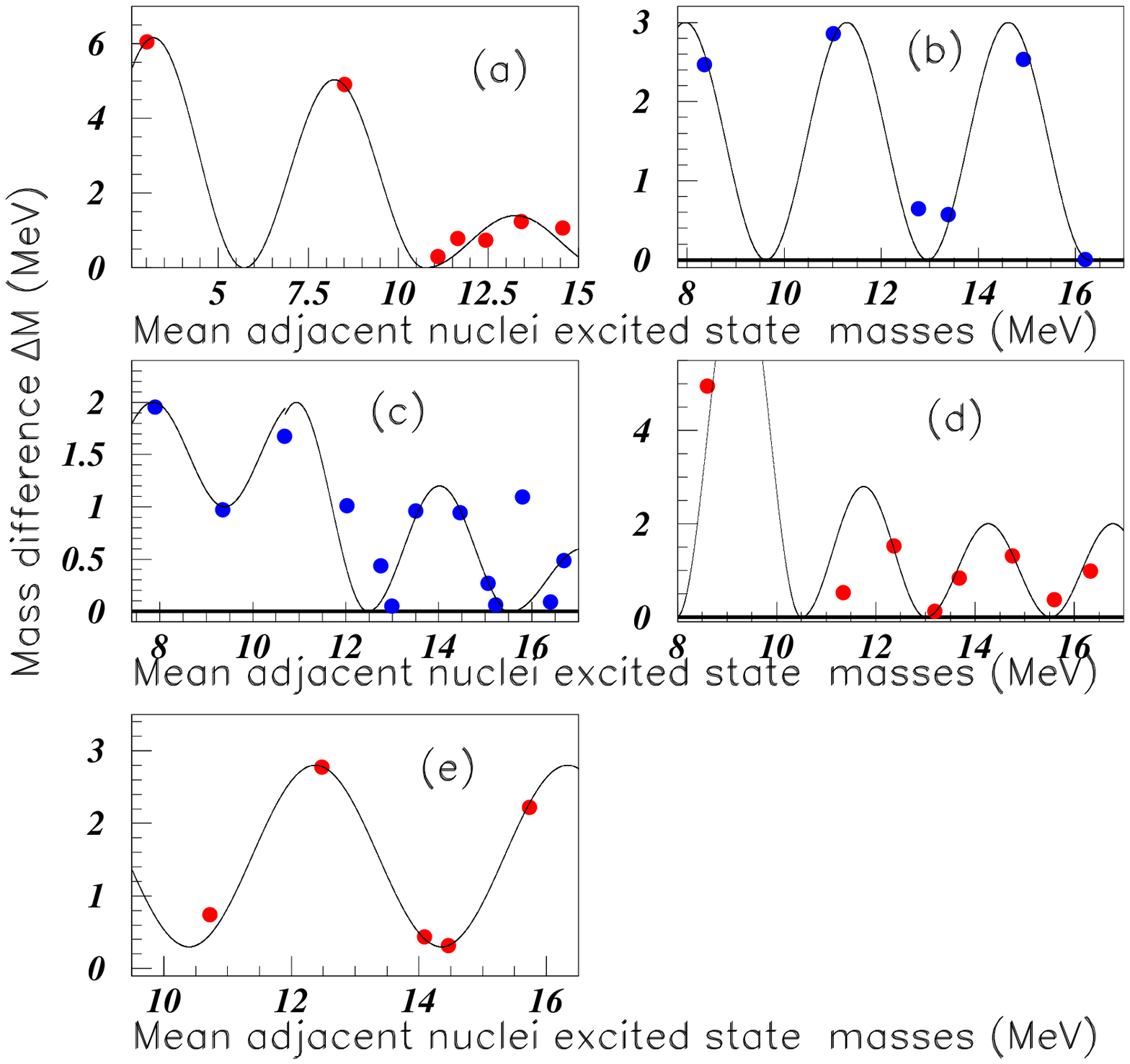}}
\caption{Color on line. Inserts (a), (b), (c), (d), and (e) show successively the mass difference between successive masses, plotted versus both corresponding mean masses for $^{16}$O nucleus: J=0, J=1, J=2, J=3, and J=4 respectively.}
\end{figure}

\begin{figure}[h]
\centering
\hspace*{-0.2cm}
\scalebox{0.58}[0.78]{
\includegraphics[bb=24 130 516 545,clip,scale=0.8] {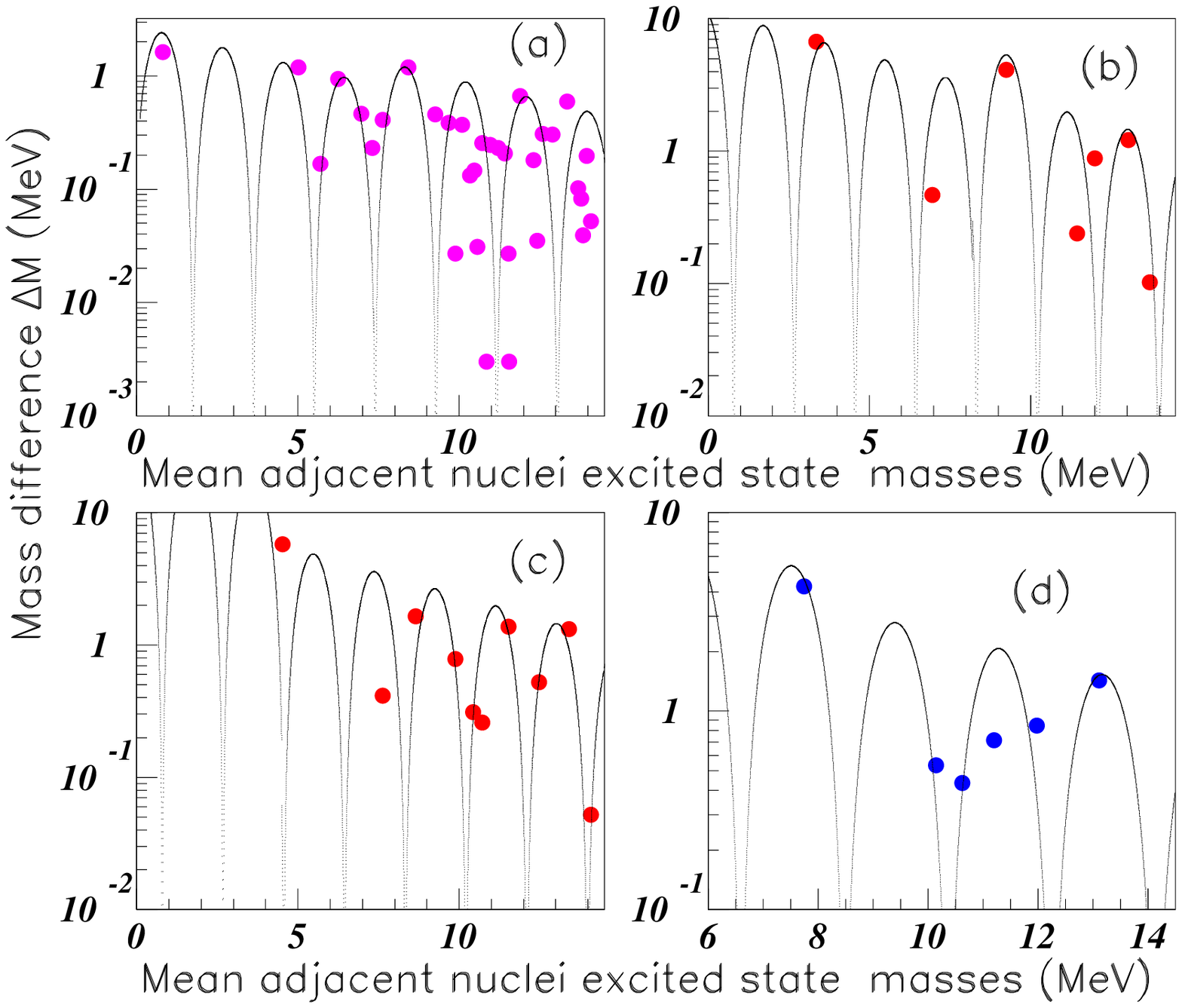}}
\caption{Color on line. Inserts (a), (b), (c), and (d) show successively the mass difference between successive masses, plotted versus both corresponding mean masses for $^{20}$Ne nucleus: all spins,  J=0, J=2, and J=3 respectively.}
\end{figure}

\begin{figure}[h]
\centering
\hspace*{-0.2cm}
\scalebox{0.58}[0.68]{
\includegraphics[bb=29 229 520 543,clip,scale=0.8] {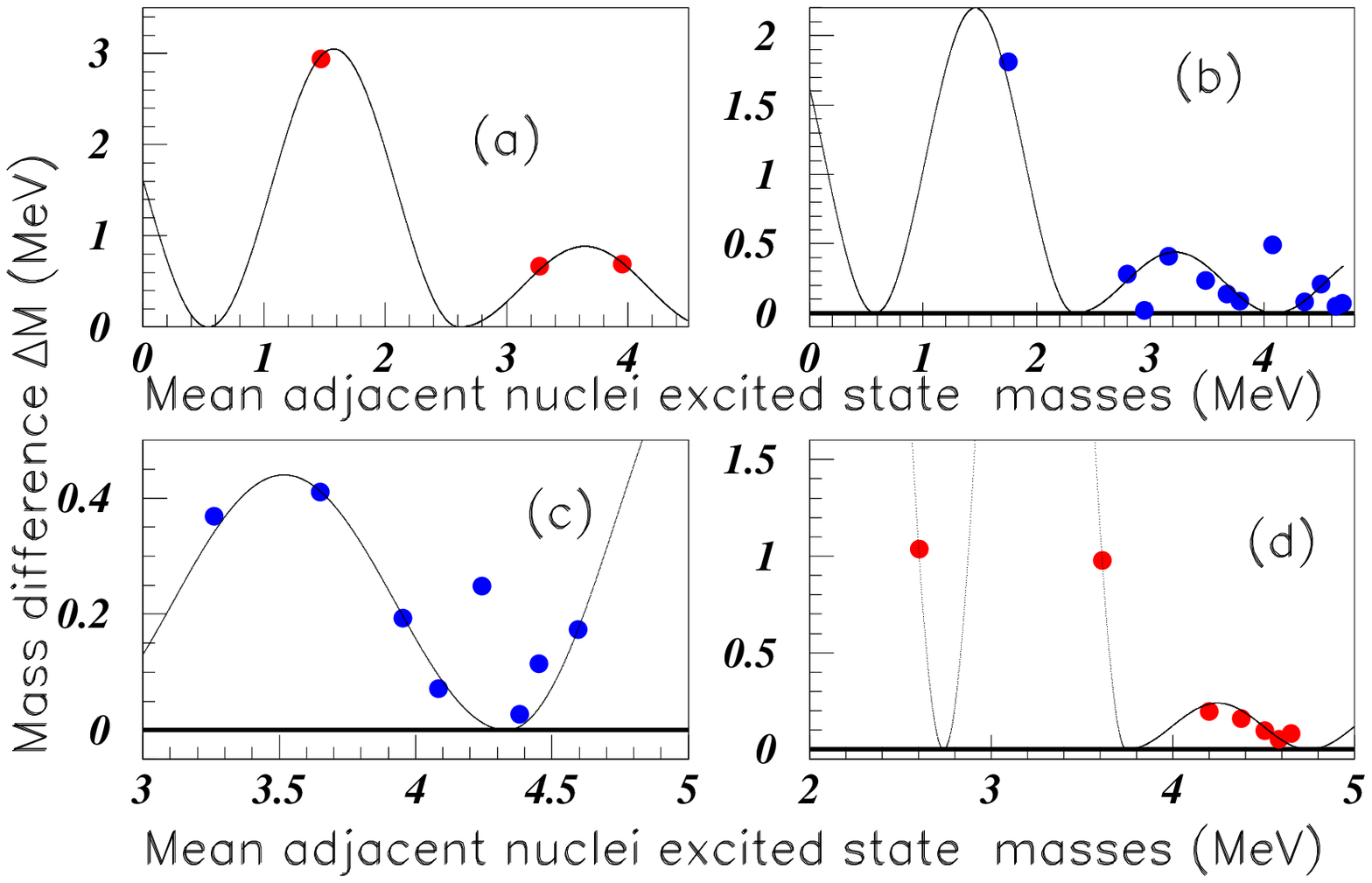}}
\caption{Color on line. Inserts (a), (b), (c), and (d) show successively the mass difference between successive masses, plotted versus both corresponding mean masses for $^{56}$Fe nucleus: J=0, J=2, J=3, and J=4 respectively.}
\end{figure}

\begin{figure}[h]
\centering
\hspace*{-0.2cm}
\scalebox{0.58}[0.74]{
\includegraphics[bb=31 128 518 544,clip,scale=0.8] {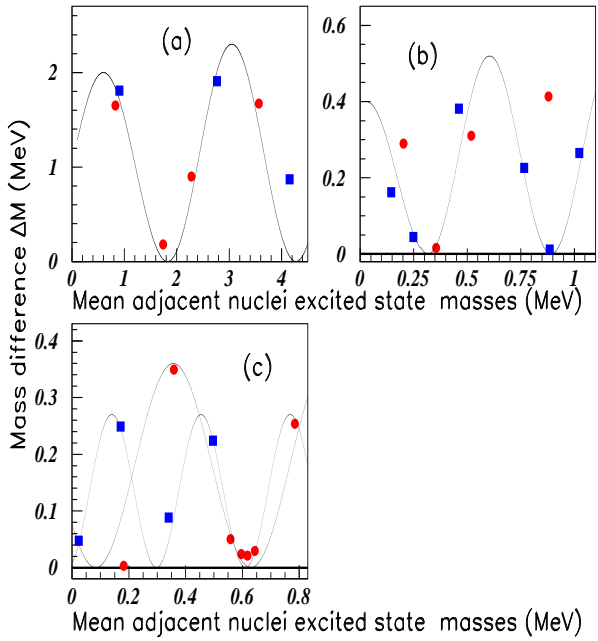}}
\caption{Color on line. Inserts (a), (b), and (c) show successively the mass difference between successive masses, plotted versus both corresponding mean masses for J=5/2  $^{25}$Al and $^{27}$Al, $^{155}$Tb and $^{159}$Tb, and $^{165}$Dy and $^{165}$Er nuclei respectively.(See text).}
\end{figure}

\begin{figure}[h]
\centering
\hspace*{-0.2cm}
\scalebox{0.58}[0.74]{
\includegraphics[bb=28 84 516 552,clip,scale=0.8] {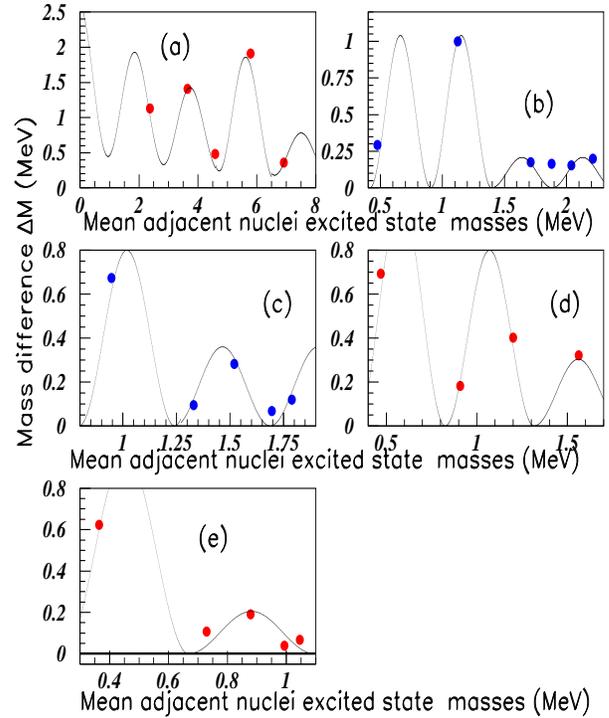}}
\caption{Color on line. Inserts (a), (b), (c), (d), and (e) show successively the mass difference between successive masses, plotted versus both corresponding mean masses for J=2  $^{26}$Mg, $^{194}$Pt, $^{214}$Po, $^{154}$Gd, and $^{230}$Th nuclei respectively.}
\end{figure}

\begin{figure}[h]
\centering
\hspace*{-0.2cm}
\scalebox{0.58}[0.74]{
\includegraphics[bb=20 82 516 546,clip,scale=0.8] {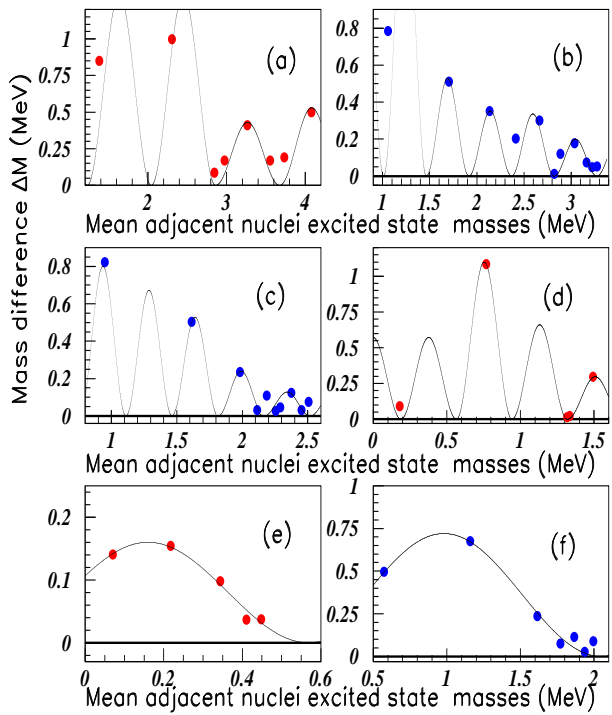}}
\caption{Color on line. Inserts (a), (b), (c), (d), (e), and (f) show successively the mass difference between successive masses, plotted versus both corresponding mean masses for J=2  $^{62}$Zn, $^{80}$Se, $^{100}$Ru, $^{92}$Nb, $^{146}$La, and $^{132}$Ce nuclei respectively.}
\end{figure}

\begin{figure}[h]
\centering
\hspace*{-0.2cm}
\scalebox{0.58}[0.74]{
\includegraphics[bb=17 239  520  545,clip,scale=0.8] {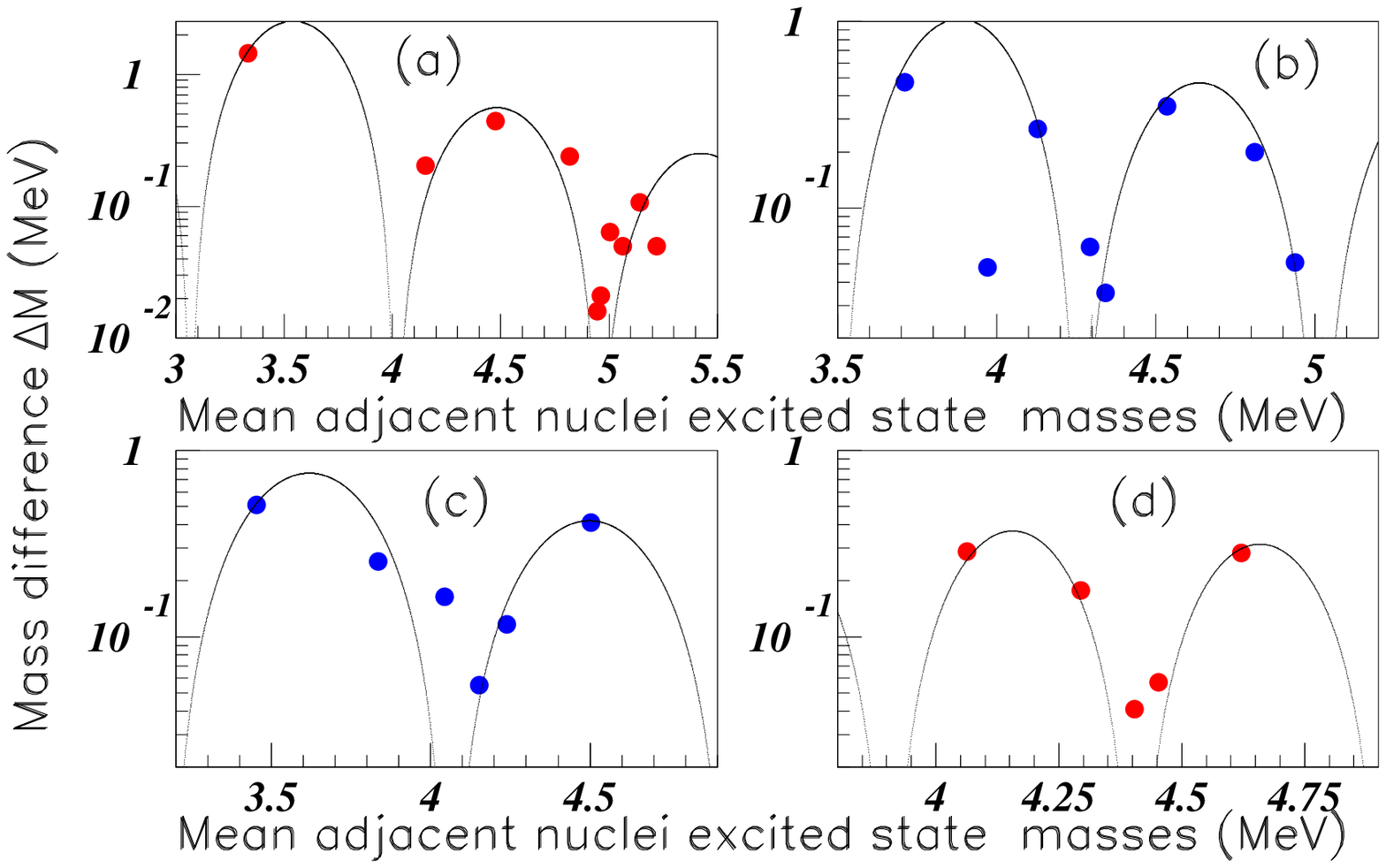}}
\caption{Color on line. Inserts (a), (b), (c), and (d) show successively the mass difference between successive masses, plotted versus both corresponding mean masses for $^{208}$Pb J=3, J=4, J=5, and J=6 respectively.}
\end{figure}

Fig.10 shows the mass difference between successive masses plotted versus the corresponding masses for $^{10}$B nucleus. The four inserts (a), (b), (c), and (d) show data with: all spins P = 1.885~MeV, J=1 P=2.2~MeV, J=2 P=2.2~MeV, and J=3 P=2.2~MeV. The data for separated spins (inserts (b), (c), and (d)) are well fitted  with the same period of oscillation. However the small number of levels with spin J=3 (insert (d)) involves an arbitrary fit.

Fig.~11 shows the mass difference between successive masses plotted versus the corresponding masses for $^{14}$N nucleus. The four inserts (a), (b), (c), and (d) show data with: all spins P = 1.82~MeV, J=1 P=2.01~MeV, J=2 P=2.01~MeV, and J=3 P=2.04~MeV. The data for separated spins (inserts (b), (c), and (d)) are well fitted  with almost the same period of oscillation.

Fig.~12 shows the mass difference between successive masses plotted versus the corresponding masses for $^{15}$N nucleus. The four inserts (a), (b), (c), and (d) show data with: all spins P = 1.885~MeV, J=1/2 P=1.885~MeV, J=3/2 P=1.885~MeV, and J=5/2 P=1.63~MeV. The data for separated spins are better fitted in inserts (b) and (d) than in (c).

Fig.~13 shows the mass difference between successive masses plotted versus the corresponding masses for $^{16}$N nucleus \cite{Tilley:1993zz}. The three inserts (a), (b), and (c) show data with: J=1 P = 1.82~MeV, J=2 P=2.07~MeV, and J=3 P=2.39~MeV. Insert (d) shows the result for the levels J=2 P=5.03~MeV of the $^{4}$He nucleus \cite{Tilley:1992zz}. The data are well fitted. 

Fig.~14 shows the mass difference between successive masses plotted versus the corresponding masses for $^{16}$O nucleus \cite{Tilley:1993zz}. The five inserts (a), (b), (c), (d), and (e) show data with: J=0 P = 5.03~MeV, J=1 P=3.33~MeV, J=2 P=3.08~MeV, J=3 P=2.51~MeV, and J=4 P=3.96~MeV. One data with   J=2, close to 16~MeV, is outside the fit. For all nuclei, at large excitation energy, the spins of some levels are unknown, therefore these levels are ignored.

Fig.~15 shows the mass difference between successive masses plotted in log scale versus the corresponding masses for $^{20}$Ne nucleus. The four inserts (a), (b), (c), and (d) show data with: all spins P = 1.885~MeV, J=0 P=1.885~MeV, J=2 P=1.885~MeV, and J=3 P=1.885~MeV. The data for separated spins (inserts (b), (c), and (d)) are rather well fitted  with  the same period of oscillation, the same as obtained for $^{15}$N (J=3/2).

Fig.~16 shows the mass difference between successive masses plotted versus the corresponding masses for $^{56}$Fe nucleus \cite{56Fe}. The four inserts (a), (b), (c), and (d) show data with: J=0 P =2.07~MeV, J=2 P=1.76~MeV, J=3 P=1.63~MeV, and J=4 P=1.00~MeV. The data are well fitted; here the oscillatory periods decrease with increasing spins.  The fit in insert (a) is undetermined due to small number of data. There is two data outside the fit in insert (b) and one in insert (c).

Fig.17 shows the mass difference between successive masses plotted versus the corresponding masses for several nuclei with spin J=5/2. The data in insert (a) correspond to $^{25}$Al (full blue squares) and $^{27}$Al (full red circles) fitted with the period P=2.45~MeV.
The data in insert (b) correspond to $^{155}$Tb (full blue squares) and $^{159}$Tb (full red circles) fitted with the period P=0.575~MeV. The fit is obtained using $^{155}$Tb data. Three red data over four, corresponding to $^{159}$Tb, lie close to the same curve. The data in insert (c) correspond to $^{165}$Er (full blue squares) fit with the period P=0.314~MeV, and $^{165}$Dy (full red circles)  \cite{165Dy} fit with the period P=0.547~MeV. Both nuclei  $^{165}$Er and $^{165}$Dy differ by only by one proton (and one neutron), therefore the large difference between their oscillating periods is unclear.

Fig.~18 shows the mass difference between successive masses plotted versus the corresponding masses for several nuclei with spin J=2. The data in insert (a), (b), (c), (d), and (e) correspond to $^{26}$Mg P=1.885~MeV, $^{194}$Pt P=0.49~MeV, $^{214}$Po P=0.446~MeV, $^{154}$Gd P=0.49~MeV, and $^{230}$Th P=0.427~MeV nuclei respectively. Here again the periods decrease with increasing nuclei masses.

Fig.~19 shows the mass difference between successive masses plotted versus the corresponding masses for several other nuclei with spin J=2. The data in insert (a), (b), (c), (d), (e), and (f) correspond to $^{62}$Zn \cite{62Zn} P=0.817, $^{80}$Se \cite{80Se} P=0.452~MeV, $^{100}$Ru \cite{100Ru} P=0.352~MeV, $^{92}$Nb P=0.377~MeV, $^{146}$La P=0.817~MeV, and  $^{132}$Ce P=2.14~MeV nuclei respectively. 

Fig.~20 shows the mass difference between successive masses plotted versus the corresponding masses for $^{208}$Pb \cite{208Pb} in log scale. Inserts (a), (b), (c), and (d) correspond respectively to data having the following spins: J=3 P=0.942~MeV, J=4 P=0.754~MeV, J=5 P=0.88~MeV, and J=6 P=0.503~MeV. 
The extracted periods do not fullfil the trend observed previously, namely to decrease with increasing spins. However only J=4 or preferably J=5 data are concerned with this comment. We observe that both corresponding inserts (b) and (c)  exhibit one data outside the fit which remains eventually doubtful, asking eventually for more data.

\begin{table}[t]
\caption{Quantitative information concerning the oscillation behavior of some J=2 nuclei levels analysed previously. P is the period (in MeV).}
\label{Table 2}
\vspace{5.mm}
\begin{tabular}{c c c}
\hline\underline{}
nucleus&fig.&P(MeV)\\
\hline
$^{4}$He&13(d)&5.03\\
$^{10}$B&10(c)&2.2\\
$^{14}$N&11(c)&2.0\\
$^{16}$N&13(b)&2.07\\
$^{16}$O&14(c)&3.08\\
$^{20}$Ne&15(c)&1.885\\
$^{26}$Mg&18(a)&1.885\\
$^{56}$Fe&16(b)&1.76\\
$^{62}$Zn&19(a)&0.817\\
$^{80}$Se&19(b)&0.452\\
$^{92}$Nb&19(d)&0.377\\
$^{100}$Ru&19(c)&0.352\\
$^{132}$Ce&19(f)&2.14\\
$^{146}$La&19(e)&0.817\\
$^{154}$Gd&18(d)&0.490\\
$^{194}$Pt&18(b)&0.490\\
$^{214}$Po&18(c)&0.446\\
$^{230}$Th&18(e)&0.427\\
\hline
\end{tabular}
\end{table}
\begin{figure}[h]
\centering
\hspace*{-0.2cm}
\scalebox{1.2}[1]{
\includegraphics[bb=36 342 295 543,clip,scale=0.8] {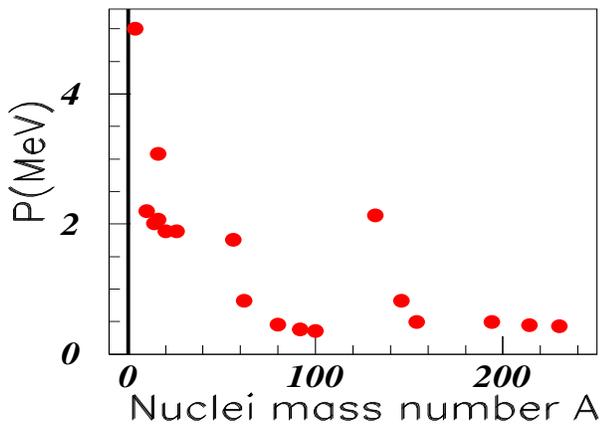}}
\caption{Color on line. Period variation versus the mass number A for J=2 levels.}
\end{figure}
\begin{table}[t]
\caption{Quantitative information concerning the oscillation behavior of some nuclei levels with spin different from 2, analysed previously. P(m) is the period (in MeV).}
\label{Table 2}
\vspace{5.mm}
\begin{tabular}{c c c c}
\hline\underline{}
Spin&nucleus&fig.&P(MeV)\\
\hline
0&$^{16}$O&14(a)&5.03\\
&$^{20}$Ne&15(b)&1.885\\
 &$^{56}$Fe&16(a)&2.07\\
\hline 
1/2&$^{15}$N&12(b)&1.885\\
\hline
1&$^{10}$B&10(b)&2.2\\ 
 &$^{14}$N&11(b)&2.01\\
 & $^{16}$N&13(a)&1.82\\
 & $^{16}$O&14(b)&3.33\\
\hline
3/2&$^{15}$N&12(c)&1.885\\
\hline 
5/2&$^{15}$N&12(d)&1.633\\
     &$^{25}$Al&17(a)&2.45\\
     &$^{27}$Al&17(a)&2.45\\
     &$^{155}$Tb&17(b)&0.575\\     
     &$^{159}$Tb&17(b)&0.575\\        
     &$^{165}$Dy&17(c)&0.547\\    
     &$^{165}$Er&17(c)&0.314\\    
\hline 
3&$^{10}$B&10(d)&2.2\\ 
 &$^{14}$N&11(d)&2.04\\
 &$^{16}$N&13(c)&2.39\\
 &$^{16}$O&14(d)&2.51\\
 &$^{20}$Ne&15(d)&1.885\\
 &$^{56}$Fe&16(c)&1.63\\
  &$^{208}$Pb&20(a)&0.94\\
\hline
4&$^{16}$O&14(e)&3.96\\
 &$^{56}$Fe&16(d)&1.005\\ 
  &$^{208}$Pb&20(b)&0.754\\ 
\hline
5&$^{208}$Pb&20(c)&0.88\\  
\hline
6&$^{208}$Pb&20(d)&0.50\\   
\hline
\end{tabular}
\end{table}

\begin{figure}
\centering
\scalebox{1.2}[1]{
\includegraphics[bb=41 331 282 546,clip,scale=0.8] {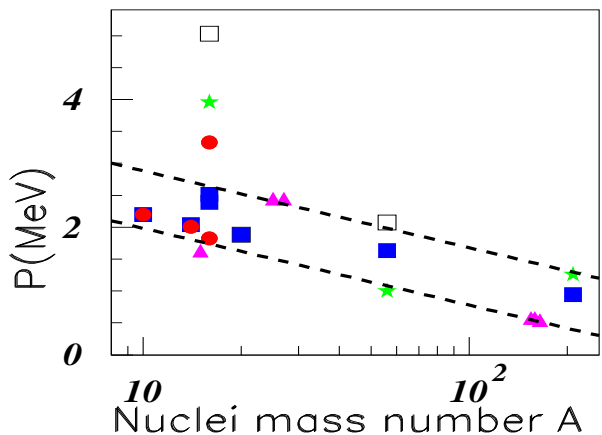}}
\caption{Color on line. Period variation versus the mass number A for J $\not=$ 2. J=1 data are shown with red full circles, J=3 with blue full squares, and J=4 with green full stars.}
\end{figure}
\section{Discussion and Conclusion}
Table II shows the periods of oscillation of the nuclei levels J=2 studied previously. Their variation versus the mass number A is displayed in fig.~21. Fast increases are observed for $^{4}$He, $^{16}$O, and $^{132}$Ce, with an abrupt fall between $^{56}$Fe and $^{62}$Zn. The two first high data are related to doubly magic number nuclei. We observe larger periods for closed shells or subshells, followed by smaller periods. The fall between $^{56}$Fe and $^{62}$Zn
should then be attributed to the passing through the magic number Z=28. The $^{132}$Ce has 58 protons which  close the ${\it 1f_{7/2}}$ subshell. 

 The missing of enough known spins for the nuclei levels with neutron (or proton) numbers close to other  magic numbers, prevents to study these mass regions. 

Table III  and fig.~22 show the periods of oscillation of the other nuclei level periods studied previously. J=0 data are shown with black empty squares, J=1 with red full circles, 
J=3 with blue full squares, J=4 with green full stars, and J=5/2 with purple full triangles. The periods concentrate between both dashed lines and decrease with increasing masses,  with a fast jump for all $^{16}$O periods J=0, 1, 3, and 4.  Table III shows that the period of variation of different spin levels remain constant for light nuclei like $^{10}$B, $^{14}$N, and $^{20}$Ne. This is also the case for $^{16}$O except for the J = 4 levels. For heavier nuclei, the periods decrease for increasing spins. 
For $^{56}$Fe nucleus for increasing spins  J=0, 2, 3, and 4, the periods are respectively: P=2.07, 1.8, 1.63, and 1.005~MeV. For $^{208}$Pb, the period for J=5 is somewhat larger than expected for a regular decrease.

When these studies considered all spins \cite{Tatischeff:2016psc}, the figures of period variation versus the masses exhibited nice shapes in agreement with a clear oscillation for the first several masses only.  In the present study done for given spins, the agreement of data versus calculations is good in all ranges where almost all levels have a known spin. 

This study should be extended for higher mass hadrons, not known presently.  It was already mentioned that a minimum of five masses of  given quantum numbers must exist. 

The same remark holds for nuclear levels. Whereas a lot of nuclear levels is known, the spin of many of them is ignored. Moreover,  there are  few levels with unknown spin, which masses are located between the masses of known spin levels. This may then alter the data of higher mass levels.

In conclusion the paper shows that the oscillating periods of mesons and baryons follow the same variation. This symmetry of oscillation is observed for the masses of hadrons and masses of nuclei levels which display  oscillatory behaviours well observed using the relation (1) and well fitted with the cosine function (2). Such behaviour requires the need for a theoretical study to describe the oscillating distributions and particularly the oscillation amplitudes.

\end{document}